\documentstyle[12pt]{article}
\title{\bf Nonperturbative Field Correlators\\
in the Abelian Higgs Model}
\author{D.V. ANTONOV \thanks{Phone: 0049-30-2093 7974; Fax: 0049-30-2093 
7631; 
E-mail address: 
antonov@pha2.physik.hu-berlin.de}{\,}
\thanks{On leave of absence from the Institute of Theoretical and 
Experimental Physics, B. Cheremushkinskaya 25, 117 218, Moscow, 
Russia.}
\\
{\it Institut f\"ur Physik, Humboldt-Universit\"at zu Berlin,}\\
{\it Invalidenstrasse 110, D-10115, Berlin, Germany}}
\date{}
\begin{document}
\maketitle
\vspace{1mm}
\centerline{\bf {Abstract}}
\vspace{3mm}
By making use of the duality transformation, gauge 
field correlators of the Abelian Higgs Model 
are studied in the London limit. 
The obtained results are  
in a good agreement 
with the dual Meissner scenario of confinement and with the Stochastic 
Model of QCD vacuum. 
The nontrivial contribution to the quartic 
correlator arising due to accounting for the finiteness of the 
coupling constant is discussed.

\vspace{6mm}
{\large \bf 1. Introduction}
\vspace{3mm}   

It is commonly argued that on the phenomenological level,  
quark confinement in QCD could be explained in 
terms of the dual Meissner effect [1] (for a recent review see [2]). 
That is why it looks 
reasonable to study various properties of the Abelian Higgs 
Model (AHM) [3-8] (see Ref. [9] for a review), 
where this effect displays itself, as a simple 
model of confinement. In this way, in Ref. [4] the duality 
transformation proposed in Ref. [3] has been applied to the lattice 
version of AHM in the London limit 
in order to reformulate its partition function 
in terms of the integral over world-sheets of the Abrikosov-Nielsen-Olesen 
strings. Then the same reformulation has been performed in the continuum 
limit in Refs. [7] and [8] for AHM with and without monopoles, 
respectively.  

In the present paper, we shall apply the duality transformation to 
the derivation of the string representations for the generating 
functionals 
of correlators of the gauge field strength tensors and of the Higgs 
currents in the London limit of AHM. By virtue of the obtained 
representations, we shall then get the bilocal correlator of the 
dual field strength tensors. It will be first 
obtained in Section 2 in a direct way, 
i.e. from the generating functional for the strength tensors correlators, 
and then in Section 3 from the correlator of two Higgs currents, by 
making use of the equations of motion. The asymptotic behaviours of the 
obtained correlator occur to be in a good agreement with the present 
lattice data concerning such a correlator in QCD [10]. The latter one 
plays an 
essential role in the Stochastic Vacuum Model (SVM) of QCD [11] 
(for a review see Refs. [2], [12], and [13]). 
This result supports 
the original conjecture of 't Hooft and Mandelstam concerning the dual 
Meissner mechanism of confinement. 

In Section 4, we shall consider AHM 
in the vicinity of the London limit, i.e. take into account the 
effects bringing about by the finiteness of the coupling constant. 
In this way, we shall see that these effects give a nontrivial 
contribution to the quartic correlator of the field strength tensors. 

The main results of the paper will be discussed in the 
Conclusion.

In the Appendix, we perform a duality transformation of the generating 
functional for the field strengthes in the London limit of AHM.

\vspace{6mm}
{\large\bf 2. String Representation for the Generating Functional  
of the Correlators of the Gauge Field Strengthes}
\vspace{3mm}

We shall start with the following expression for the generating 
functional  
of the dual gauge field strength tensors in AHM

$${\cal Z}\left[S_{\alpha\beta}\right]=
\int \left|\Phi\right| {\cal D}\left|
\Phi\right| {\cal D}A_\mu {\cal D}\theta^{{\rm sing.}}
{\cal D}\theta^{{\rm reg.}}
\exp\Biggl\{-\int d^4x\Biggl[\frac14 F_{\mu\nu}^2+\frac12\left|D_\mu\Phi
\right|^2+\lambda\left(\left|\Phi\right|^2-\eta^2\right)^2+$$

$$+iS_{\mu\nu}\tilde F_{\mu\nu}\Biggr]
\Biggr\}, \eqno (1)$$
where $\Phi(x)=\left|\Phi(x)\right| {\rm e}^{i\theta(x)}$ is the 
Higgs field, $\theta=
\theta^{{\rm sing.}}+\theta^{{\rm reg.}}$,  
$D_\mu\equiv\partial_\mu-ieA_\mu$ is the covariant derivative, and 
$S_{\alpha\beta}$ is a source term. 

It is worth mentioning from the 
very beginning, that in what follows we shall be interested in the 
correlators of the {\it dual} field strengthes, rather than in the 
ordinary ones. This is because our main goal will be the derivation 
of the two coefficient functions $D$ and $D_1$, which parametrize 
the bilocal field strength correlator in the SVM (see Eqs. (12) and 
(13) below). Should we parametrize in such a way the correlator of 
the {\it usual} field strengthes, then as it has been explained 
in Refs. [2] and [12], the function $D$ is nonvanishing only in the 
case of AHM with monopoles (otherwise, the term with the function 
$D$ violates Abelian Bianchi identities). However, this problem is 
absent for the bilocal correlator of the {\it dual} field strengthes, 
in which the function $D$ is nontrivial and by virtue of the 
equations of motion could be found from the correlator of two Higgs 
currents (see the next Section).

In the London  
limit, $\lambda\to\infty$, one has $\left|\Phi\right|\to\eta$, and 
Eq. (1) takes the form

$${\cal Z}\left[S_{\alpha\beta}\right]=
\int {\cal D}A_\mu {\cal D}\theta^{{\rm sing.}}
{\cal D}\theta^{{\rm reg.}}\exp 
\left\{-\int d^4x\left[\frac14 F_{\mu\nu}^2+\frac{\eta^2}{2}\left(
\partial_\mu\theta-eA_\mu\right)^2+iS_{\mu\nu}\tilde F_{\mu\nu}
\right]\right\}. \eqno (2)$$

Performing   
the duality transformation of Eq. (2) along 
the lines described in Refs. [3] and [5-7], we get 

$${\cal Z}\left[S_{\alpha\beta}\right]=
\int {\cal D}A_\mu \int {\cal D}h_{\mu\nu}\int {\cal D}x_\mu(\xi) 
\exp\left\{\int d^4x\left[-\frac1{12\eta^2}H_{\mu\nu
\lambda}^2+i\pi h_{\mu\nu}\Sigma_{\mu\nu}-\frac14 F_{\mu\nu}^2-\right.
\right.$$

$$\left.\left.-i\tilde F_{\mu\nu}\left(\frac{e}{2}h_{\mu\nu}+S_{\mu\nu}
\right)\right]\right\}, \eqno (3)$$
where $\Sigma_{\mu\nu}(x)\equiv\int
\limits_\Sigma^{}d\sigma_{\mu\nu}(x(\xi))\delta(x-x(\xi))$ is the 
vorticity tensor current defined on the world-sheet $\Sigma$ of the closed 
Abrikosov-Nielsen-Olesen string, $\xi=\left(\xi^1, \xi^2\right)$ is a 
two-dimensional coordinate, and  
$H_{\mu\nu\lambda}\equiv\partial_\mu h_{\nu\lambda}+
\partial_\lambda h_{\mu\nu}+\partial_\nu h_{\lambda\mu}$ is a field 
strength 
tensor of an antisymmetric tensor field $h_{\mu\nu}$. 
The details of derivation of Eq. (3) are presented in the Appendix.

Now it is convenient to rewrite 

$$\exp\left(-\frac14\int d^4x 
F_{\mu\nu}^2\right)=
\int {\cal D} G_{\mu\nu}\exp\left\{\int d^4x
\left[-G_{\mu\nu}^2+i\tilde F_{\mu\nu}
G_{\mu\nu}\right]\right\},$$
after which $A_\mu$-integration yields 

$$\int {\cal D}A_\mu \exp\left\{-\int d^4x\left[\frac14 F_{\mu\nu}^2+
i\tilde F_{\mu\nu}\left(\frac{e}{2}h_{\mu\nu}+S_{\mu\nu}\right)
\right]\right\}=$$

$$=\int {\cal D}G_{\mu\nu}\exp\left(-\int d^4x G_{\mu\nu}^2
\right)\delta\left(\varepsilon_{\mu\nu\lambda\rho}\partial_\mu
\left(G_{\lambda\rho}-\frac{e}{2}h_{\lambda\rho}-S_{\lambda\rho}\right)
\right)=$$

$$=\int {\cal D}B_\mu\exp\left[\int d^4x\left(\frac{e}{2}h_{\mu\nu}
+S_{\mu\nu}+\partial_\mu B_\nu-\partial_\nu B_\mu\right)^2\right]. 
\eqno (4)$$

Next, performing the hypergauge transformation with the function $\frac{2}
{e}B_\mu$ (see e.g. Ref. [5]), we get from Eq. (4) 

$${\cal Z}\left[S_{\alpha\beta}\right]=
\exp\left(-\int d^4x S_{\mu\nu}^2\right)\cdot$$

$$\cdot\int {\cal D}x_\mu(\xi)
\int {\cal D}
h_{\mu\nu}\exp \left\{\int d^4x\left[-\frac{1}{12\eta^2}H_{\mu\nu
\lambda}^2
-\frac{e^2}{4}h_{\mu\nu}^2+ih_{\mu\nu}\left(\pi\Sigma_{\mu\nu}
+ieS_{\mu\nu}\right)\right]\right\}.\eqno (5)$$

Gaussian 
integration over the field $h_{\mu\nu}$ in Eq. (5) 
leads to the following final result for the string representation 
of the generating functional (1) 

$${\cal Z}\left[S_{\alpha\beta}\right]=
\exp\left(-\int d^4x S_{\mu\nu}^2\right)\cdot$$

$$\cdot\int {\cal D}x_\mu(\xi)\exp\left\{-\int d^4x\int d^4y
\left(\pi\Sigma_{\lambda\nu}(x)+ieS_{\lambda\nu}(x)\right)D_{\lambda\nu, 
\mu\rho}(x-y)\left(\pi\Sigma_{\mu\rho}(y)+ieS_{\mu\rho}(y)\right)
\right\}. \eqno (6)$$

It is worth mentioning, that an analogous expression for the generating 
functional has been for the first time obtained in terms of 
another fields already in the last 
paper in Ref. [1] and further studied in the first three papers in 
Ref. [2]. There it has been done by making use of the techniques quite 
different from the ones applied in the present paper, i.e. without 
casting the generating functional into the form of the integral 
over the Kalb-Ramond field. However, the final results presented 
in these papers qualitatively correspond to our Eq. (6).  

In Eq. (6), the propagator of the field $h_{\mu\nu}$ has the following form

$$D_{\lambda\nu, \mu\rho}(x)\equiv D_{\lambda\nu, \mu\rho}^{(1)}(x)+
D_{\lambda\nu, \mu\rho}^{(2)}(x),$$
where

$$D_{\lambda\nu, \mu\rho}^{(1)}(x)=\frac{e\eta^3}{8\pi^2}\frac
{K_1}{\left|x\right|}\Biggl(\delta_{\lambda\mu}\delta_{\nu\rho}-
\delta_{\mu\nu}\delta_{\lambda\rho}\Biggr), \eqno (7)$$

$$D_{\lambda\nu, \mu\rho}^{(2)}(x)=\frac{\eta}{4\pi^2ex^2}\Biggl[\Biggl(
\frac{K_1}{\left|x\right|}+\frac m2\left(K_0+K_2\right)\Biggr)
\Biggl(\delta_{\lambda\mu}\delta_{\nu\rho}-\delta_{\mu\nu}\delta_{\lambda
\rho}\Biggr)+$$

$$+\frac{1}{2\left|x\right|}\Biggl[3\Biggl(\frac{m^2}{4}
+\frac{1}{x^2}\Biggr)K_1+\frac{3m}{2\left|x\right|}\left(K_0+
K_2\right)+\frac{m^2}{4}K_3\Biggr]\cdot$$

$$\cdot\Biggl(\delta_{\lambda\rho}x_\mu x_\nu+\delta_{\mu\nu}x_\lambda 
x_\rho-\delta_{\mu\lambda}x_\nu x_\rho-\delta_{\nu\rho}x_\mu x_\lambda
\Biggr)\Biggr]. \eqno (8)$$
Here $K_i\equiv K_i(m\left|x\right|), i=0,1,2,3,$ stand for the modified 
Bessel functions, and $m\equiv e\eta$ is the mass of the gauge boson 
generated by the Higgs mechanism. It is worth noting that the term 

$$\int\limits_\Sigma^{} 
d\sigma_{\lambda\nu}(x)\int\limits_\Sigma^{} 
d\sigma_{\mu\rho}(y)D_{\lambda\nu, 
\mu\rho}^{(2)}(x-y)$$
could be rewritten as a boundary one, and therefore vanishes, since 
$\Sigma$ is 
closed. 

Let us now derive from the general form (6) of the generating functional 
for the correlators of the field strengthes the expression for the bilocal 
correlator. The result reads

$$\left. \left<\tilde F_{\lambda\nu}(x)\tilde F_{\mu\rho}(y)\right>=
\frac{1}{{\cal Z}[0]}\frac{\delta^2{\cal Z}\left[S_{\alpha\beta}\right]}
{\delta S_{\lambda\nu}(x)\delta S_{\mu\rho}(y)}
\right|_{S_{\alpha\beta}=0}=$$

$$=\left(\delta_{\lambda\mu}\delta_{\nu\rho}-\delta_{\lambda\rho}
\delta_{\mu\nu}\right)\delta(x-y)+
2e^2\Biggl[D_{\lambda\nu, \mu\rho}(x-y)-$$

$$-2\pi^2
\left<\int\limits_\Sigma^{}
d\sigma_{\alpha\beta}(z)\int\limits_\Sigma^{}d\sigma_{\gamma\zeta}(u)
D_{\alpha\beta, \mu\rho}(z-y)D_{\gamma\zeta, \lambda\nu}(u-x)
\right>_{x_\mu(\xi)}\Biggr], \eqno (9)$$
where

$$\left<...\right>_{x_\mu(\xi)}\equiv\frac{\int {\cal D}x_\mu(\xi)(...)
\exp\left[-\pi^2\int\limits_\Sigma^{}d\sigma_{\alpha\beta}(z)
\int\limits_\Sigma^{}d\sigma_{\gamma\zeta}(u)D_{\alpha\beta, \gamma\zeta}
(z-u)\right]}{\int {\cal D}x_\mu(\xi)
\exp\left[-\pi^2\int\limits_\Sigma^{}d\sigma_{\alpha\beta}(z)
\int\limits_\Sigma^{}d\sigma_{\gamma\zeta}(u)D_{\alpha\beta, \gamma\zeta}
(z-u)\right]},$$
and the term with the $\delta$-function 
on the R.H.S. of Eq. (9) corresponds to the free contribution to the 
correlator. 
According to Eqs. (7) and (8), one can see that due to the large distance 
asymptotics 
of the modified Bessel functions, $D_{\lambda\nu, \mu\rho}(x)$ has 
the order of magnitude $e^2\eta^4$. Therefore, the contribution of 
the second term in the square brackets on the R.H.S. 
of Eq. (9) to the bilocal correlator of the dual gauge 
field strengthes is much smaller than the contribution arising from 
the first term, when 

$$e\eta^2\left|\Sigma\right|\ll 1, \eqno (10)$$ 
where 
$\left|\Sigma\right|$ stands for the area of the surface 
$\Sigma$. Such an inequality holds at least in the weak coupling 
regime (which is in the line with the London limit) 
or in the case of sufficiently small $\eta$.
 
Following the SVM, let us parametrize the bilocal 
correlator of the dual field strengthes 
by the two Lorentz structures as follows

$$\left<\tilde F_{\lambda\nu}(x)\tilde F_{\mu\rho}(0)\right>=
\Biggl(\delta_{\lambda\mu}\delta_{\nu\rho}-\delta_{\lambda\rho}
\delta_{\nu\mu}\Biggr)D\left(x^2\right)+$$

$$+\frac12\Biggl[\partial_\lambda
\Biggl(x_\mu\delta_{\nu\rho}-x_\rho\delta_{\nu\mu}\Biggr)
+\partial_\nu\Biggl(x_\rho\delta_{\lambda\mu}-x_\mu\delta_{\lambda\rho}
\Biggr)\Biggr]D_1\left(x^2\right). \eqno (11)$$
In Eq. (11), the division into such Lorentz structures 
is chosen in such a way that one can due to the Stokes theorem rewrite 
the double surface integral of the structure 
with the function $D_1$ as a boundary one, whereas this is impossible 
to do for the structure with the function $D$. Then in the approximation 
(10), by virtue of Eqs. (7) and (8), we arrive at the following values 
of the functions $D$ and $D_1$

$$D\left(x^2\right)=\frac{m^3}{4\pi^2}\frac{K_1}{\left|x\right|} 
\eqno (12)$$
and 

$$D_1\left(x^2\right)=\frac{m}{2\pi^2x^2}\Biggl[\frac{K_1}{\left|x\right|}
+\frac{m}{2}\Biggl(K_0+K_2\Biggr)\Biggr]. \eqno (13)$$
In Eq. (12), we have neglected the free $\delta$-function type contribution.
The asymptotic behaviours of the coefficient functions (12) and (13) at 
$\left|x\right|\ll\frac1m$ and $\left|
x\right|\gg\frac1m$ read 

$$D\longrightarrow\frac{m^2}{4\pi^2x^2}, \eqno (14)$$

$$D_1\longrightarrow\frac{1}{\pi^2\left|x\right|^4} \eqno (15)$$
and

$$D\longrightarrow\frac{m^4}{4\sqrt{2}\pi^{\frac32}}
\frac{{\rm e}^{-m\left|x\right|}}{\left(m\left|x\right|\right)^
{\frac32}}, \eqno (16)$$

$$D_1\longrightarrow\frac{m^4}{2\sqrt{2}\pi^{\frac32}}
\frac{{\rm e}^{-m\left|x\right|}}{\left(m\left|x\right|\right)^
{\frac52}}, \eqno (17)$$
respectively. 

One can now see that according to the lattice data [10] 
the asymptotic behaviours (14)-(17) are very similar 
to the ones of the functions $D$ and $D_1$, which parametrize the 
gauge-invariant bilocal correlator of gluonic field strengthes in 
the SVM approach to QCD. In particular, at large distances both 
functions decay exponentially, and the function $D$ is much larger 
then the function $D_1$, whereas at small distances the 
function $D_1$ behaves as $\frac{1}{\left|x\right|^4}$ and is 
much larger than the function $D$ in this limit, which is in the line 
with the SVM. This similarity in the large- and short distance 
asymptotic behaviours of the functions $D$ and $D_1$, which 
parametrize the bilocal correlator of the dual field strengthes 
in AHM and the gauge-invariant correlator in QCD, 
supports the original conjecture 
by 't Hooft and Mandelstam concerning the dual Meissner nature of 
confinement.  

\vspace{6mm}
{\large \bf 3. String Representation for the Generating Functional  
of the Higgs Currents Correlators}
\vspace{3mm}

In this Section, we shall rederive the coefficient function $D$ in the 
bilocal correlator of the 
dual field strength tensors 
from the string representation for the 
generating functional of correlators of the Higgs currents. In the 
London limit, such a generating functional reads

$$\hat {\cal Z}\left[J_\mu\right]=
\int {\cal D}A_\mu {\cal D}\theta^{{\rm sing.}}{\cal D}\theta^{{\rm reg.}}
\exp
\left\{\int d^4x\left[-\frac14 F_{\mu\nu}^2-\frac{\eta^2}{2}\left(
\partial_\mu\theta-eA_\mu\right)^2+J_\mu j_\mu\right]\right\},$$
where $j_\mu\equiv -e\eta^2(\partial_\mu\theta-eA_\mu)$ is the 
Higgs current in this limit. 

Performing the same duality transformation as the one used in 
Section 2, we get the following string representation for $\hat {\cal Z}
\left[J_\mu\right]$

$$\hat {\cal Z}\left[J_\mu\right]=\exp\left[
\frac{m^2}{2}\int d^4x J^2(x)\right]
\int {\cal D}x_\mu (\xi)
\exp\left[-\pi^2\int\limits_\Sigma^{}d\sigma_{\alpha\beta}(z)
\int\limits_\Sigma^{}d\sigma_{\gamma\zeta}(u)D_{\alpha\beta, \gamma\zeta}
(z-u)\right]\cdot$$

$$\cdot\exp\Biggl\{e\varepsilon_{\lambda\nu\alpha\beta}
\int d^4x d^4y\Biggl[
-\frac{e}{4}\varepsilon_{\mu\rho\gamma\delta}\Biggl(\frac{\partial^2}
{\partial x_\alpha\partial y_\gamma}D_{\lambda\nu, \mu\rho}(x-y)
\Biggr)J_\beta (x) J_\delta (y)+$$

$$+\pi \Sigma_{\mu\rho}(y)\Biggl(
\frac{\partial}{\partial x_\alpha} D_{\lambda\nu, \mu\rho}(x-y)\Biggr)
J_\beta (x)\Biggr]\Biggr\}. \eqno (18)$$
Varying Eq. (18) twice w.r.t. $J_\mu$, setting then $J_\mu$ equal to zero, 
and dividing the result by $\hat {\cal Z}[0]$, we arrive 
at the following expression for the correlator of two Higgs currents  

$$\left<j_\beta(x) j_\sigma(y)\right>=m^2\delta_{\beta\sigma}\delta(x-y)+
e^2 
\varepsilon_{\lambda\nu\alpha\beta}\varepsilon_
{\mu\rho\gamma\sigma}\Biggl[-\frac12\frac{\partial^2}{\partial 
x_\alpha \partial y_\gamma}D_{\lambda\nu, \mu\rho}(x-y)+$$

$$+\pi^2\left< 
\int\limits_\Sigma^{} d\sigma_{\delta\zeta}(z)\int\limits_\Sigma^{} 
d\sigma_{\chi\varphi}(u)
\Biggl(\frac{\partial}{\partial x_\alpha}D_{\lambda\nu, 
\delta\zeta}(x-z)\Biggr)\Biggl(\frac{\partial}
{\partial y_\gamma} D_{\mu\rho, \chi\varphi}(y-u)\Biggr)
\right>_{x_\mu(\xi)}\Biggr]. \eqno (19)$$

It is  easy to see that the contribution of the 
term (8) to the R.H.S. of Eq. (19) vanishes, whereas the contribution of 
Eq. (7) to the second term in the square brackets on the R.H.S. 
of Eq. (19) could be disregarded w.r.t. 
its contributon to the first term, if the inequality (10) holds. 
Within this 
approximation, making use of the 
equation\footnote{An analogous equation for the dual fields has been 
first presented in Ref. [2]. For the first time, the fact that the 
function $D$ becomes nonvanishing in QED with magnetic monopoles 
has been mentioned in Ref. [12].}

$$\left<j_\beta(x)j_\sigma(y)\right>=\Biggl(\frac{\partial^2}
{\partial x_\mu\partial y_\mu}\delta_{\beta\sigma}-
\frac{\partial^2}{\partial x_\beta\partial y_\sigma}\Biggr)
D\left((x-y)^2\right), \eqno (20)$$
which follows from the equation of motion, 
we get from Eqs. (19), (20), and (7) the following value of the function $D$

$$D\left(x^2\right)=\frac{m^3}{4\pi^2}\frac{K_1
\left(m\left|x\right|\right)}
{\left|x\right|}, \eqno (21)$$
where we have not again presented the free contribution.  
One can see that Eq. (21) coincides with the value of the function $D$ given 
by Eq. (12) obtained from the string representation for the partition 
function of correlators of the dual gauge field strengthes. 
Notice also, that only 
the function $D$ could be obtained from the correlator (19) 
due to the 
independence of the latter of the function $D_1$.  

\vspace{6mm}
{\large \bf 4. $1/{\lambda}$-Corrections to the 
Quartic Correlator of the Dual Field Strength Tensors}
\vspace{3mm}

In this Section, we shall demonstrate that accounting for the finiteness 
of the coupling constant $\lambda$ in AHM leads to a nontrivial 
contribution 
to the quartic correlator of the dual field strength tensors. 
Let us notice from the very beginning that we shall not be interested 
in free contributions to the correlators and therefore in what follows 
consider only the correlators of the fields defined in different 
space-time points. 
In 
the London limit, within the 
approximation (10), the dominant contribution to the correlator 

$$\left<\tilde F_{\mu_1\nu_1}(x_1)\tilde F_{\mu_2\nu_2}(x_2)
\tilde F_{\mu_3\nu_3}(x_3)\tilde F_{\mu_4\nu_4}(x_4)\right>,~ 
x_1\ne x_2\ne x_3\ne x_4,$$
comes from the four-fold variation of the following term in the 
expansion of the generating functional (6)

$$\frac{e^4}{2}\int d^4xd^4yd^4zd^4u S_{\mu\nu}(x)D_{\mu\nu, 
\lambda\rho}(x-y)S_{\lambda\rho}(y) S_{\alpha\beta}(z)D_{\alpha\beta, 
\gamma\zeta}(z-u)S_{\gamma\zeta}(u). \eqno (22)$$
The $4!=24$ terms containing 
all possible combinations of indices and arguments, which one gets during 
this variation, are obvious and we shall not list them here for 
shortness.  

Let us now consider AHM in the vicinity of the London limit. To this 
end, we shall expand the radial part of the Higgs field, $\left|
\Phi(x)\right|$, in Eq. (1) as $\left|
\Phi(x)\right|=\eta+\tau\psi(x)$, where $\tau\equiv\frac{1}{\lambda}\to 
0$, and $\psi(x)$ is an arbitrary quantum fluctuation. Correspondingly, 
in what follows we shall keep only the terms linear in $\tau$. Then 
neglecting the Jacobian emerging during the change 
of the integration variables, $\left|\Phi(x)\right|\to\psi(x)$, 
which will be eventually cancelled after division of the final 
expression for the field correlator by ${\cal Z}[0]$, we 
arrive at the new expression for the generating functional (1), which 
reads 

$${\cal Z}\left[S_{\alpha\beta}\right]=\int {\cal D}\psi {\cal D}
A_\mu {\cal D}\theta^{\rm sing.}{\cal D}\theta^{\rm reg.}\exp\Biggl\{
-\int d^4x\Biggl[\frac14 F_{\mu\nu}^2+\frac{\tau^2}{2}(\partial_\mu 
\psi)^2+4\tau\eta^2\psi^2+$$

$$+\eta\Biggl(\frac{\eta}{2}+\tau\psi\Biggr)(\partial_\mu\theta-
eA_\mu)^2+iS_{\mu\nu}\tilde F_{\mu\nu}\Biggr]\Biggr\}. \eqno (23)$$
Following the same steps which led from Eq. (1) to Eq. (6), we get 
from Eq. (23) an additional weight factor in the integral 
over string world-sheets standing in Eq. (6). This weight factor 
emerges due to the 
$\psi$-integration and reads 

$$\exp\Biggl\{\frac{M}{288\pi^2\eta^6}\int d^4xd^4y 
H_{\mu\nu\lambda}^{{\rm extr.}{\,}2}(x)\frac{K_1\left(M\left|x-y
\right|\right)}{\left|x-y\right|} 
H_{\alpha\beta\gamma}^{{\rm extr.}{\,}2}(y)\Biggr\}. \eqno (24)$$
In Eq. (24), $M\equiv 2\sqrt{2\lambda}\eta$ is the mass of the Higgs 
field, and $H_{\mu\nu\lambda}^{\rm extr.}$ stands 
for the strength tensor of the saddle-point value of the field 
$h_{\mu\nu}$ following from Eq. (5). 
Taking into account that 
$\partial_\mu\Sigma_{\mu\nu}=0$, since $\Sigma$ is a closed surface, 
we get the following result for this saddle-point value 

$$h_{\lambda\nu}^{\rm extr.}(x)=\frac{m}{2\pi^2}\int d^4y
\frac{K_1(m\left|x-y\right|)}{\left|x-y\right|}\Biggl\{\eta^2
\left[i\pi\Sigma_{\lambda
\nu}(y)-eS_{\lambda\nu}(y)\right]+$$

$$+\frac{1}{e}\partial_\rho\left[\partial_\nu S_{\lambda\rho}(y)-
\partial_\lambda S_{\nu\rho}(y)\right]\Biggr\}. \eqno (25)$$

Let us prove that the terms with the derivatives of $S_{\mu\nu}$ on the 
R.H.S. of Eq. (25) yield zero. Due to the Hodge decomposition theorem 
(see e.g. [14]), $S_{\mu\nu}$ could be always represented in the form 
$S_{\mu\nu}=\partial_\mu K_\nu-\partial_\nu K_\mu+
\partial_\alpha L_{\alpha\mu\nu}$, where $K_\mu$ and $L_{\alpha\mu\nu}$ 
stand for some vector and an antisymmetric rank-3 tensor respectively. 
In the product $S_{\mu\nu}\tilde F_{\mu\nu}$, the contribution of 
the vector $K_\mu$ obviously vanishes due to the partial integration, 
and we are left with the 
representation of $S_{\mu\nu}$ of the form $S_{\mu\nu}=
\partial_\alpha L_{\alpha\mu\nu}$. Substituting it into Eq. (25) we 
see that the terms with the derivatives of $S_{\mu\nu}$ 
on the R.H.S. of this equation vanish. 

Finally, upon substitution of the rest of Eq. (25) into Eq. (24), we 
arrive at the following additional term 
in the expansion of the 
generating functional, whose contribution to the quartic correlator 
in the approximation (10) is dominant 
  
$$\frac{e^2Mm^2}{32\pi^2}\int d^4xd^4yd^4ud^4vd^4zd^4w\frac{K_1\left(
M\left|x-y\right|\right)}{\left|x-y\right|}D_1\left((x-u)^2\right)
D_1\left((x-v)^2\right)D_1\left((y-z)^2\right)\cdot$$

$$\cdot D_1\left((y-w)^2\right)
(x-u)_\mu (y-z)_\alpha\Biggl[(x-v)_\mu (y-w)_\alpha S_{\nu\lambda}(u)
S_{\nu\lambda}(v)S_{\beta\gamma}(z)S_{\beta\gamma}(w)+$$

$$+4(x-v)_\lambda (y-w)_\gamma S_{\nu\lambda}(u)S_{\mu\nu}(v)
S_{\beta\gamma}(z)S_{\alpha\beta}(w)+$$

$$+4(x-v)_\mu (y-w)_\gamma S_{\nu\lambda}(u)S_{\nu\lambda}(v)
S_{\beta\gamma}(z)S_{\alpha\beta}(w)\Biggr]. \eqno (26)$$ 
In Eq. (26), $D_1$ stands for the coefficient function (13) entering 
the bilocal correlator in the London limit. Even without explicit 
writing down $3\cdot 4!=72$ terms following from Eq. (26) after its 
four-fold 
variation, 
we see that the resulting quartic correlator differs from the one 
of the London limit, which could be obtained from Eq. (22). The most 
crucial difference is due to the presence of the Higgs boson 
exchange in Eq. (26), which is absent in Eq. (22). 

The other important outcome of Eq. (26) is that the leading correction to 
the quartic correlator, which arises due to the finiteness of $\lambda$,  
could be {\it completely} described in terms of the coefficient 
function (13), if the inequality (10) holds.

\vspace{6mm}
{\large \bf 5. Conclusion}
\vspace{3mm}

In the present paper, we have studied dual gauge field strength 
correlators in AHM 
both in the London limit and beyond. To this end in Section 2, the 
generating functional for these correlators in the London limit has been 
cast into the form (6) of the integral over world-sheets of the 
Abrikosov-Nielsen-Olesen strings. In approximation (10), this 
generating functional 
yielded the values of the two coefficient functions,  
$D$ and $D_1$, which parametrize the bilocal correlator 
and play a key role in SVM of QCD. These functions 
are given by Eqs. (12) and (13), and there asymptotic behaviours 
(14)-(17) at small and large distances are found to be in a good 
agreement with the known lattice data [10] concerning the corresponding 
behaviours in QCD. This fact together with SVM 
supports the conjecture of 't Hooft and 
Mandelstam about the dual Meissner nature of confinement. In Section 3, 
by making use of the equations of motion, 
we have rederived the coefficient function $D$ from the string 
representation for the correlator of two Higgs currents 
(19). In Section 4, we have studied AHM in the vicinity of the 
London limit and demonstrated that there appears a nontrivial contribution 
to the quartic correlator of the dual field strengthes. This contribution 
is due to accounting for the finiteness of the Higgs boson mass, which 
leads to intermediate exchanges by this boson in the quartic correlator. 
Besides that, we have demonstrated that according to Eq. (26), in 
approximation (10), the quartic correlator near the London limit 
could be described only in terms of the function $D_1$, given by Eq. 
(13).

\vspace{6mm}
{\large \bf 6. Acknowledgments}
\vspace{3mm}

The author is deeply grateful to Drs. N. Brambilla, M.N. Chernodub, 
H. Dorn, Chr. Preitschopf, A. Vairo, and 
especially to Profs. 
H.G. Dosch, M.I. Polikarpov, and Yu.A. Simonov for 
useful discussions. He would also like to thank Prof. H. Kleinert for 
bringing his attention to some papers from Refs. [1] and [2],  
the theory group of the 
Quantum Field Theory Department of the Institute of Physics of the 
Humboldt University of Berlin for kind hospitality, and Graduiertenkolleg 
{\it Elementarteilchenphysik} for financial support. 

\vspace{6mm}
{\large \bf Appendix. Derivation of Eq. (3).}
\vspace{3mm}

In this Appendix, we shall present some details of derivation of Eq. (3). 
Firstly, one can linearize the term $\left(\partial_\mu\theta-eA_\mu
\right)^2$ on the R.H.S. of Eq. (2) and carry out the integral over 
$\theta^{{\rm reg.}}$ as follows 

$$\int {\cal D}\theta^{{\rm reg.}}\exp\left\{-\frac{\eta^2}{2}
\int d^4x \left(\partial_\mu\theta-eA_\mu
\right)^2\right\}=$$

$$=\int {\cal D}C_\mu {\cal D}\theta^{{\rm reg.}}
\exp\left\{\int d^4x\left[-\frac{1}{2\eta^2}C_\mu^2+iC_\mu
\left(\partial_\mu\theta-eA_\mu
\right)\right]\right\}=$$

$$=\int {\cal D}C_\mu\delta\left(\partial_\mu C_\mu\right)
\exp\left\{\int d^4x\left[-\frac{1}{2\eta^2}C_\mu^2+iC_\mu
\left(\partial_\mu\theta^{{\rm sing.}}-eA_\mu
\right)\right]\right\}. \eqno (A.1)$$
The constraint $\partial_\mu C_\mu=0$ could be uniquely resolved by 
representing $C_\mu$ in the form $C_\mu=\frac12
\varepsilon_{\mu\nu\lambda\rho}\partial_\nu h_{\lambda\rho}$, where 
$h_{\lambda\rho}$ stands for an antisymmetric tensor field. 
Notice, that the number 
of degrees of freedom during such a replacement 
is conserved, since both of the fields 
$C_\mu$ and $h_{\mu\nu}$ have three independent components.

Now, taking 
into account that $\varepsilon_{\mu\nu\lambda\rho}\partial_\lambda
\partial_\rho\theta^{{\rm sing.}}(x)=2\pi\Sigma_{\mu\nu}(x)$ [3], we get 
from Eq. (A.1) 

$$\int {\cal D}\theta^{{\rm sing.}}
{\cal D}\theta^{{\rm reg.}}\exp\left\{-\frac{\eta^2}{2}
\int d^4x \left(\partial_\mu\theta-eA_\mu
\right)^2\right\}=$$

$$=\int {\cal D}h_{\mu\nu} {\cal D}x_\mu (\xi)\exp\left\{\int d^4x 
\left[-\frac{1}{12\eta^2}H_{\mu\nu\lambda}^2+i\pi h_{\mu\nu}
\Sigma_{\mu\nu}-\frac{ie}{2}\varepsilon_{\mu\nu\lambda\rho}A_\mu
\partial_\nu h_{\lambda\rho}\right]\right\}. \eqno (A.2)$$
In derivation of Eq. (A.2), 
we have replaced ${\cal D}\theta^{{\rm sing.}}$ by 
${\cal D}x_\mu(\xi)$ (since the surface $\Sigma$ parametrized by 
$x_\mu(\xi)$ is just the surface, at which the field 
$\theta$ is singular) and for 
simplicity have not taken into account the Jacobian arising during 
such a change of the integration variable \footnote{For the case when 
the surface $\Sigma$ has a spherical topology, this Jacobian 
has been calculated in Ref. [7].}. 

Finally, adding the term $-iS_{\mu\nu}\tilde F_{\mu\nu}$ to the argument 
of the exponent standing on the R.H.S. of Eq. (A.2), we arrive at 
Eq. (3).

\vspace{6mm}
{\large \bf References}

\vspace{3mm}
\noindent
$[1]$~S. Mandelstam, Phys. Lett. 
B 53 (1975) 476; 
G. 't Hooft, in: High Energy Physics, ed. A. Zichichi  
(Italy, Bologna, 1976); H. Kleinert, Gauge Fields in Condensed Matter, 
Vol. 1: Superflow and Vortex Lines. Disorder Fields, Phase Transitions 
(World Scientific Publishing Co., Singapore, 1989), Phys. Lett. B 293 
(1992) 168.\\
$[2]$~M. Kiometzis, H. Kleinert, and A.M.J. Schakel, Phys. Rev. Lett. 73 
(1994) 1975, Fortschr. Phys. 43 (1995) 697; H. Kleinert, Theory of 
Fluctuating Nonholonomic Fields and Applications: Statistical Mechanics 
of Vortices and Defects and New Physical Laws in Spaces with Curvature 
and Torsion, in: Proceedings of NATO Advanced Study Institute on 
Formation and Interactions of Topological Defects at the University 
of Cambridge, England, Plenum Press, New York, 1995 (preprint 
cond-mat/9503030 (1995));
Yu.A. Simonov, Phys. Usp. 
39 (1996) 313.\\
$[3]$~K. Lee, Phys. Rev. D 48 (1993) 2493.\\
$[4]$~M.I. Polikarpov, 
U.-J. Wiese, and M.A. Zubkov, Phys. Lett. B 309 
(1993) 133.\\ 
$[5]$~P. Orland, Nucl. Phys. B 428 (1994) 221.\\
$[6]$~M. Sato and S. Yahikozawa, Nucl. Phys. B 436 (1995) 100.\\
$[7]$~E.T. Akhmedov, M.N. Chernodub, M.I. Polikarpov, and M.A. Zubkov, 
Phys. Rev. 
D 53 (1996) 2087.\\ 
$[8]$~E.T. Akhmedov, JETP Lett.  
64 (1996) 82.\\ 
$[9]$~M.N. Chernodub and M.I. Polikarpov, preprints hep-th/9710205 
and ITEP-TH-55-97 (1997).\\ 
$[10]$A. Di Giacomo and H. Panagopoulos, Phys. Lett. B 285  
(1992) 133; L. Del Debbio, A. Di Giacomo, and Yu.A. Simonov, 
Phys. Lett. B 332 (1994) 111; M. D'Elia, A. Di Giacomo, and 
E. Meggiolaro, Phys. Lett. B 408 (1997) 315; A. Di Giacomo, 
E. Meggiolaro, and H. Panagopoulos, Nucl. Phys. B 483 (1997) 371;
A. Di Giacomo, E. Meggiolaro, and H. Panagopoulos, Nucl. Phys. 
Proc. Suppl. A 54 (1997) 343.\\
$[11]$H.G. Dosch, Phys. Lett. 190 (1987) 177; Yu.A. Simonov, 
Nucl. Phys. B 307 (1988) 512; H.G. Dosch and Yu.A. Simonov, Phys. Lett.  
B 205 (1988) 339; 
H.G. Dosch and E. Ferreira, Phys. Lett. B 318 (1993) 197; 
H.G. Dosch, H.J. Pirner, and Yu.A. Simonov, Phys. Lett. B 349 (1995) 335; 
H.G. Dosch et al., Phys. Lett. B 365 (1996) 213,  
Phys. Rev. D 55 (1997) 2602, Nucl. Phys. B 507 (1997) 519; 
Yu.A. Simonov, Phys. Atom. Nucl. 60 (1997) 2069; 
H.G. Dosch, T. Gousset, and H.J. Pirner, Phys. Rev. D 57 (1998) 1666.\\
$[12]$Yu.A. Simonov, Sov. J. Nucl. Phys. 54 (1991) 115.\\
$[13]$H.G. Dosch, Prog. Part. Nucl. Phys. 33 (1994) 121.\\
$[14]$M. Nakahara, Geometry, Topology, and Physics 
(England, London, 1990).

\end{document}